\documentclass[11pt,a4paper]{article}
\usepackage{jheppub}
\usepackage{amsmath}
\usepackage{amssymb}

\newcommand{\bea}{\begin{eqnarray}}
\newcommand{\eea}{\end{eqnarray}}
\newcommand{\bit}{\begin{itemize}}
\newcommand{\eit}{\end{itemize}}

\def\nl{\nonumber \\}

\def\a{\alpha}

\def\b{\beta}

\def\s{\sigma}

\def\p{\partial}

\def\le{\left(}
\def\ri{\right)}

\def\beq{\begin{equation}}
\def\eeq{\end{equation}}

\def\arr{{\rightarrow}}

\def\hV {\hat{V}}
\def\th {\tilde{h}}

\title{Heat kernel for Newton-Cartan trace anomalies} 

\author[a,b]{Roberto Auzzi,} 
\author[a,c]{and Giuseppe Nardelli}

\affiliation[a]{Dipartimento di Matematica e Fisica, Universit\`a Cattolica
del Sacro Cuore, \\
Via Musei 41, 25121 Brescia, Italy}
\affiliation[b]{INFN Sezione di Perugia, \\ Via A. Pascoli, 06123 Perugia, Italy}
\affiliation[c]{TIFPA - INFN, c/o Dipartimento di Fisica, Universit\`a di Trento, 38123 Povo (TN), Italy}

\emailAdd{roberto.auzzi@unicatt.it}
\emailAdd{giuseppe.nardelli@unicatt.it}

\abstract{
We compute the leading part of the trace anomaly 
 for a free non-relativistic scalar in $2+1$ dimensions
coupled to a background Newton-Cartan metric.
The anomaly is proportional to $1/m$,
where $m$ is the mass of the scalar. We comment on the
implications of a conjectured $a$-theorem for non-relativistic 
theories with boost invariance.
}


\begin{document}

\maketitle


 \section{Introduction}
\label{sec_intro}

Conformal anomalies have a long and glorious history in quantum field theory,
see e.g. \cite{Duff:1977ay,Duff:1993wm}. 
For relativistic quantum field theories in even dimension,
they are a very useful tool to characterize
the irreversibility properties of the Renormalization Group (RG).
In $2$ dimensions this is established by Zamolodchikov's 
$c$-theorem  \cite{Zamolodchikov:1986gt}: in this case it is possible to build
a monotonically decreasing quantity defined also outside the fixed points,
which coincides with the conformal anomaly at the endpoints of the RG flow. 
In $4$ dimensions a similar property (known as $a$-theorem) was conjectured by Cardy
 \cite{Cardy:1988cwa} and later established nearby
 weakly coupled fixed points in \cite{Osborn:1989td,Jack:1990eb,Osborn:1991gm}.
 Dispersion relations  of dilaton scattering amplitudes 
 \cite{Komargodski:2011vj,Komargodski:2011xv}
 were used for a non-perturbative proof.
 
Scale-invariant  fixed points are common 
not only in high energy physics,
but also in non-relativistic condensed matter systems.
It would be of general interest to establish some non-relativistic
version of the $a$-theorem, which might be used to constraint
RG flows in non-relativistic strongly coupled systems,
such as fermions at unitarity, and to classify the landscape of possible 
fixed point realized in many-body physics.

Scale anomalies are natural candidates for such 
monotonically decreasing functions along the RG flow.
In the non-relativistic case, scale invariance is
characterized by a different scaling
of the time and space coordinates.
Such a different scaling  can be parameterized by the dynamical exponent $z$:
\beq
x^i \arr e^{\s} x^i \, ,\qquad t \arr e^{z \s} t \, . 
\eeq
In all these situations, we may expect a quantum
violation of scale invariance due to coupling to background
curved spacetime:
\beq
T^i_i -z \epsilon^0 = \mathcal{A} \, ,
\eeq
where $T^i_i$ and $\epsilon^0$ are the spatial
stress-tensor components and the energy density,
and $\mathcal{A}$ a function of the background curvatures
and gauge fields.

In order to study these issues, first of all one needs to couple the 
field theory which one is studying to a background curved spacetime.
The kind of background depends on the symmetries of the theory.
In particular, a very different anomaly structure is found
depending if we require or not non-relativistic boost invariance.

In the case without boosts,
studied in  \cite{Adam:2009gq,Baggio:2011ha,Griffin:2011xs,Arav:2014goa,Arav:2016xjc},
several anomalies are indeed possible at the scale-invariant fixed points.
Unfortunately, in all the cases that have been studied, these anomalies
have vanishing Weyl variation (type B anomalies \cite{Deser:1993yx}).
Consequently, an analysis based on the Wess-Zumino
consistency conditions as the one 
in  \cite{Osborn:1989td,Jack:1990eb,Osborn:1991gm}
would not give any constraint on the RG flow of the anomaly
coefficients.

 The case with boost invariance instead looks much more promising.
 In this case it is natural to couple the non-relativistic
 theory to a Newton-Cartan (NC) gravity background.
 The study of the anomaly
 in two spatial dimensions and for $z=2$
 was initiated in \cite{Jensen:2014hqa}.
 The outcome was that in this case, by dimensional analysis,  an infinite
 number of terms  is in principle possible in the anomaly.
 Moreover, there is a selection rule which splits 
 these  terms into distinct sectors, each
 with a finite numbers of terms and 
 decoupled Wess-Zumino consistency conditions.
 In the simplest sector, it turns out that 
 the anomaly structure is exactly the same as
 for a relativistic theory in four spacetime dimensions.
 
In order for a  NC background 
to be consistent with causality, the Frobenius condition
$n \wedge dn=0$ should be satisfied, $n_\mu $ being a nowhere-vanishing
 1-form identifying the local time direction.
It turns out that the structure of the anomaly critically depends 
whether causal backgrounds are or not allowed.
If backgrounds which do not satisfy the Frobenius condition
are discarded, the structure of the anomaly becomes much simpler;
in particular there is just a finite number of terms in the anomaly \cite{Auzzi:2015fgg}.
 Moreover, in a rather subtle way,  the type A anomaly disappears
 \cite{Arav:2016xjc,Auzzi:2015fgg} and just a type B one survives.
 
 One may worry about the possibility that
  coupling the theory to a non-causal background could be 
  logically inconsistent. On the other hand, technically
the NC gravity background is introduced just as a source
for the components of the non-relativistic energy-momentum (EM) tensor.
When we make a functional derivative with respect to the background fields,
indeed we do not restrict just to casual backgrounds, otherwise
we would not be able to get all the independent components
of the EM tensor.   In other words, functional sources in the path integral
are not restricted to any physical condition. Consequently,
 an unconstrained
not necessarily causal background should be used for 
the purpose of studying anomalies.

In this paper, we compute the NC trace anomalies 
with the heat kernel method
in the case of a Schr\"odinger-invariant 
 free non-relativistic
scalar in $2$ spatial dimensions.
We find that the anomaly is given by:
\beq
\mathcal{A}=
\left( -a \, E_4 + c \, W^2 + b \, R^2 + d \, D^2 R  \right) + \dots \, ,
\label{mainresult}
\eeq
where
\beq
a=\frac{1}{8 m \pi^2} \frac{1}{360}  \, , \quad
c=\frac{1}{8 m \pi^2} \frac{3}{360} \, , \quad
b=\frac{1}{8 m \pi^2} \frac{1}{2} \le \xi -\frac{1}{6}\ri^2 \, , \quad
d=\frac{1}{8 m \pi^2}  \frac{1-5 \xi}{30} \, .
\label{mainresult2}
\eeq
It is important to stress that
 the quantities $(E_4,W^2,R^2,D^2 R)$
 in eq.~(\ref{mainresult})
are completely determined in terms of the 3-dimensional
Newton-Cartan gravity fields and 
do not have anything of 4-dimensional.
Technically, $R$, $E_4$, $W^2$, $D_A$ are defined as the
scalar curvature, the Euler density, the Weyl tensor squared
and the covariant derivative of 
the extra-dimensional null reduction
in eq.~(\ref{DLCQ01}), but this is just a trick
to build quantities which are automatically
invariant under the galilean boost symmetry.
The dots in eq.~(\ref{mainresult}) correspond to 
possible terms with a higher number of derivatives.
The parameter  $\xi$ is the 
coupling of the scalar to the null reduction curvature 
$R \phi \phi^\dagger$ (conformal coupling is achieved for $\xi=1/6$).
The coefficient $d$ is a scheme-dependent quantity \cite{Bonora:1985cq},
while  $b$ vanishes for the conformal coupling.
The coefficients $a$ and $c$ correspond to genuine 
scheme-independent anomalies.
In particular, $a$ is the coefficient of a type A anomaly
and then a good candidate for a quantity which decreases along the RG flow.
As far as we know, this is the first explicit calculation of the trace anomaly 
in the Schr\"odinger case.
Our calculation is genuinely $2+1$ dimensional
and does not make use of the extra-dimensional null reduction.

In section \ref{sec2} we introduce the notation,
the coupling of the scalar to the background geometry,
and the form of the anomaly nearby flat spacetime.
In section \ref{sec3} we compute the anomaly 
using the heat kernel method.
We conclude in section \ref{sec4}.

\section{Preliminaries}

\label{sec2}

\subsection{Newton-Cartan gravity}

A NC geometry in $d+1$ spacetime dimensions is defined 
by a  1-form $n_\mu$ (which corresponds to the local time direction),
by a positive-definite symmetric tensor $h^{\mu \nu}$ with rank $d$
for which $n_\mu$ is a zero eigenvector 
\beq n_\mu h^{\mu \a}=0 \, ,
\label{corta1}
\eeq
and by a background gauge field $A_\mu$ for the particle number symmetry.
A vector field $v^\mu$, whose projection onto $n_\mu$ is one 
\beq
n_\mu v^\mu =1 \, ,
\label{corta2}
\eeq
 is also introduced; once $v^\mu$ is fixed,
it is possible to uniquely define  a degenerate rank $d$ symmetric tensor $h_{\mu \nu}$,
which corresponds to the metric along the spatial directions, 
which satisfies:
\beq
h^{\mu \a}h_{\a \nu}=\delta^\mu_\nu -v^\mu n_\nu=P^\mu_\nu \, , \qquad
h_{\mu \a} v^\a =0 \, ,
 \eeq
 where $P^\mu_\nu$ is the projector onto the spatial directions.
The NC geometry was first introduced as a
tool to write newtonian gravity in a
 diffeomorphism-invariant fashion; for a review see  \cite{gravitation}.
Recently it was realized in \cite{Son:2005rv,Hoyos:2011ez,Son:2013rqa,Geracie:2014nka}
that it is a very useful tool for condense-matter physics,
 because it is a very convenient way to parameterize the sources
of the non-relativistic energy-momentum tensor.
Other uses and applications have been recently discussed
in several papers, e.g.~\cite{Brauner:2014jaa,Jensen:2014aia,Jensen:2014wha,
Christensen:2013lma,hartong,Hartong:2014pma,Moroz:2015jla}. 

Among the symmetries of the NC theory,
besides diffeomorphisms and local $U(1)$ gauge symmetry, 
there is also a local version of the galilean boost symmetry,
which is called Milne boost.
If we denote by $\psi_\mu$ the local boost parameter,
the geometry fields transform in the following way:
\bea
v'^\mu & = & v^\mu+h^{\mu \nu} \psi_\nu \, \nl
h'_{\mu \nu} & = & h_{\mu \nu} -(n_\mu P_\nu^\rho+ n_\nu P_\mu^\rho) \psi_\rho
+n_\mu n_\nu h^{\rho \sigma} \psi_\rho \psi_\sigma \, , \nl
A'_\mu & = & A_\mu+P^\rho_\mu \psi_\rho -\frac{1}{2} n_\mu h^{\a \b} \psi_\a \psi_\b \, , 
\eea
while $n_\mu$ and $h^{\mu \nu}$ are invariant.

These non-trivial transformation properties
render  the classification of local invariant quantities complicated.
For this reason it is convenient to use 
an extra-dimensional null reduction $(x^-,x^\mu)$
from a relativistic parent space \cite{Duval:1984cj}:
\bea
\label{DLCQ01}
G_{MN} & = &\left(
\begin{array}{cc}
 0& n_\mu \\ n_\nu \ \  & n_\mu A_\nu + n_\nu A_\mu + h_{\mu \nu} \\
  \end{array}\right) =
  \left(
\begin{array}{cc}
 0& n_\mu \\ n_\nu  &  (h_A)_{\mu \nu} \\
  \end{array}\right)\, , 
      \nl
  G^{MN} & = &
  \left(
\begin{array}{cc}
 A^2-2 v \cdot A \ \ & v^\mu - h^{\mu \sigma} A_\sigma \\ 
v^\nu - h^{\nu \sigma} A_\sigma  &  h^{\mu \nu} \\
  \end{array}\right) =
    \left(
\begin{array}{cc}
\phi_A & v_A^\mu \\ 
v_A^\mu &  h^{\mu \nu} \\
  \end{array}\right) 
 \, ,
\eea
where the quantities $h_A$, $\phi_A$  and $v_A$ (which are Milne boost invariants)
are introduced. Diffeomorphism-invariant quantities in $d+2$ dimensions
are automatically Milne boost-invariant in the non-relativistic $d+1$ dimensional theory.
We will sometimes refer to this null reduction  as DLCQ, Discrete Light-Cone Quantization.
We denote by $D_A$ the covariant derivative defined by the 
Levi-Civita connection from the metric in eq.~(\ref{DLCQ01}).
It is important to stress that, even if we are often using this extra-dimensional trick,
we will compute the anomaly of the  non-relativistic theory in $d+1$ spacetime dimensions, 
and not of the $d+2$ dimensional relativistic parent theory.

It is useful to introduce the DLCQ vector
\beq
n^M=(1,0, \dots) \, , \qquad n_{M}=(0,n_\mu) \, 
\eeq
which is a null killing vector of the metric~(\ref{DLCQ01}).

A Weyl transformation on the NC background
is equivalent to a Weyl transformation on the DLCQ background
which is independent from the null direction $x^-$:
\beq
n^A D_A \s =0 \, .
\eeq
The Weyl transformation parameter $\s$ 
 is an arbitrary function of $x^\mu$;
the transformation laws of the basic  metric objects is as follows:
\beq
G_{MN} \arr e^{2 \s} G_{MN} \, , \qquad
n_\mu \arr e^{2 \s} n_\mu \, , \qquad
h_{\mu \nu} \arr e^{2 \s} h_{\mu \nu}  \, .
\eeq
In order to define a spacetime volume element
we introduce:
\beq
\sqrt{g}=\sqrt{\det (n_\mu n_\nu + h_{\mu \nu})} =\sqrt{-\det G_{AB}}\, .
\eeq

\subsection{Anomalies nearby a flat backround}

Arbitrary variations on background fields are not in general allowed, because
one must satisfy both eq.~(\ref{corta1}) and eq.~(\ref{corta2}).
The most general perturbations can be parameterized in terms
of an arbitrary $\delta n_\mu$, a transverse
 perturbation $\delta u^\mu$ with
 $\delta u^\mu n_\mu =0$ and a transverse metric 
 perturbation $\delta \th^{\a \b} n_\b=0$.
 The variation of the metric fields are then:
 \beq
 \delta n_\mu \, , \qquad
 \delta v^\mu = - v^\mu v^\a \delta n_\a + \delta u^\mu \, , \qquad
 \delta h^{\mu \nu} =-v^{\mu} \delta n^{\nu}-\delta n^\mu v^\nu 
 -\delta \th^{\mu \nu} \, .
 \label{varii}
 \eeq
 Specializing eq.~(\ref{varii}) nearby the flat limit gives:
 \bea
 n_\mu  &=&  (1+\delta n_0 , \delta n_i) \, , \qquad
 v^\mu =(1-\delta n_0, \delta u_i) \, , \qquad \delta \th^{0 i}=0 \, ,
 \nl
  h_{\mu \nu} &=&
  \left(
\begin{array}{cc}
 0& -\delta u_i \\ -\delta u_i  &  \delta_{ij} + \delta \th_{ij} \\
  \end{array}\right) \, , \qquad
h^{\mu \nu}=
  \left(
\begin{array}{cc}
 0& -\delta n_i \\ -\delta n_i  &  \delta_{ij} - \delta \th_{ij} \\
  \end{array}\right) \, .
\eea 
In terms of DLCQ extra-dimensional fields, this corresponds to:
 \bea
G_{A B} & = &
  \left(
\begin{array}{ccc}
0 & 1 + \delta n_0 &  \delta n_i \\
1+\delta n_0 & 2 \delta A_0 & \delta A_i -\delta u_i \\ 
\delta n_i  & \delta A_i -\delta u_i &\delta_{ij} + \delta \th_{ij} \\
  \end{array}\right) \, ,
  \nl
  G^{A B} & = &
  \left(
\begin{array}{ccc}
-2 A_0 & 1- \delta n_0 & -\delta A_i+ \delta u_i \\
1-\delta n_0 &  0& -\delta n_i \\ 
-\delta A_i+ \delta u_i  &  -\delta n_i&\delta_{ij} + \delta \th_{ij} \\
  \end{array}\right) \, .
  \label{perturba}
\eea

We can use these sources to define conserved currents.
Let us consider the vacuum functional $W[g_{\mu \nu}]$:
\beq
e^{i W[g_{\mu \nu}]}= \int {\cal D} \phi \,  e^{i S[\phi, g_{\mu \nu}]}
\eeq
where $\phi$ runs over the dynamical fields of the theory.
We can define the expectation values of the
energy-momentum tensor multiplet through: 
\beq
\delta W = \int d^d x\sqrt{-g} \le
\frac{1}{2} T_{i j} \delta \th_{i j} + j^\mu \delta A_\mu
-\epsilon^\mu \delta n_\mu - p_i \delta u_i 
\ri \, .
\eeq
Here $p_i$ is the momentum density,
$T_{i j}$ is the spatial stress tensor, $j^\mu=(j^0,j^i)$ contains
the number density and current and $\epsilon^\mu=(\epsilon^0,\epsilon^i)$
 the energy density and current. Number current is proportional to 
the momentum density (this is direct consequence of
eq.~(\ref{perturba}), because only the combination
 $\delta A_i - \delta u_i$ enters inside the DLCQ metric).

Conservations laws  in the flat limit give:
\beq
\p_\mu j^\mu=0 \, , \qquad 
\p_\mu \epsilon^\mu=0 \, , 
\qquad
 \p_t p^j + \p_i T^{ij } =0 \, .
\eeq

If one makes a Weyl variation $\Delta$
of the vacuum function, nearby flat spacetime
 at the first order finds:
 \beq
\Delta W= \s G_{AB} \frac{\delta W}{\delta G_{AB}} = 
\s \le \delta^{ij} \frac{\delta W}{\delta (\delta \th_{ij}) } +
2 \frac{\delta W}{\delta (\delta n_0)} \ri =
\s( T^i_i -2 \epsilon^0) \, .
\eeq
The non relativistic trace anomaly 
in $d=2$ and for $z=2$
can be conveniently written \cite{Jensen:2014hqa}
in this way:
\beq
\Delta W=\int \sqrt{g} d^3 x\  \s 
\left( -a E_4 + c W^2 + b R^2 + d D_A D^A R +e R^{AB}_{\,\,\,\,\,\,\, CD} 
R_{ABEF} \frac{\epsilon^{CDEF}}{\sqrt{g}}
\right) + \dots
\label{anoma}
\eeq
In this equation the tensors  $R_{ABCD}$, $E_4$, $W^2$ are the
Riemann curvature, the Euler density and the Weyl tensor squared of the DLCQ
metric eq.~(\ref{DLCQ01}).
As in the relativistic case in $4$ dimensions, 
in this equation $a$, $c$ and $e$ correspond to anomaly coefficients,
while $b=0$ from the Wess-Zumino consistency conditions \cite{Bonora:1983ff} and 
$d$ can be removed by local counterterms.
The dots in eq.~(\ref{anoma}) correspond to an infinite number of possible
terms with a higher number of derivatives, which however belong to 
separated Weyl sectors.
These terms are obtained contracting the DLCQ vector $n_A$
with combinations of curvatures. By  dimensional analysis,
for each $n_A$ one can add one extra DLCQ derivatives $D_A$
(being a DLCQ curvature a commutator of two covariant derivatives,
 two $n_A$ are needed in order to buy a curvature).
 Examples of possible terms with the right dimension in order
 to enter the anomaly are:
 \beq
 n^A D_A R_{BC} R^{BC} \, , \qquad
 R_{ABCP} R^{ABCQ} R^P_{\, \, \, \, MQN} n^M n^N \, .
\label{anoma-esempio}
 \eeq
The number of $n_A$ vectors, which we denote by $N_n$, is unchanged by a Weyl transformation,
so the Wess-Zumino consistency conditions can be solved independently
in each sector with different $N_n$.
Eq.~(\ref{anoma}) refers to $N_n=0$, while
an analysis of the sectors of the anomaly 
with $N_n > 0$ is left as a topic for future investigation.

We can now expand the anomaly around the flat background.
We set $\delta u_i=0$, because 
nearby flat space it is equivalent to $-\delta A_i$.

In the following eqs.~(\ref{erre}-\ref{W2}) we drop the $\delta$'s
in front of the perturbations of $n_\mu$, $A_\mu$, and $\th_{ij}$.
The DLCQ scalar curvature is:
\beq
\label{erre}
R = 
- \p^2_k  \th_{ii} + \p_{ij} \th_{ij} 
 -2 \p^2_k  n_0  +2 \p_0 (\p_i n_i) \, ,
\eeq
where $\p_{ij}= \p_i\p_j$ and  $\p^2_k=\p_k \p_k$ is the spatial flat laplacian.
The Euler density $E_4$ and the Weyl tensor squared $W^2$ read:
\bea
\label{E4}
E_4 & = &2 (\p_k ( \p_k  n_0 +\p_{0} n_k))^2
-2 (\p_i( \p_{j} n_0+ \p_{ 0 } n_j) )^2
-2 (\p_0( \epsilon_{ij} \p_{i} n_j) )^2 
\nl & &
+4 \p_0 (\p_0 n_i-\p_{i} n_0) \p_k ( \p_k n_i-\p_{i} n_k )  \, ,
\eea
\bea
\label{W2}
W^2 & = &\frac{1}{3} \le 
- \p^2_k n_0 +  \p_0 (\p_i n_i) +\p^2_k \th_{ii} - \p_{i j} \th_{ij}
\ri^2 -3 (\p_0 (\epsilon_{ij} \p_i n_j) )^2
\nl & &
+2 \p_k(\epsilon_{ij} \p_i n_j) [  \p_k ( \epsilon_{lm} \p_l A_m) 
+ \epsilon_{kl} \p_0 (\p_0 n_l - \p_{l} n_0) 
-\epsilon_{lm} \p_{0l}  \th_{km} ] \, .
\eea

\subsection{Non-relativistic scalar}

The action for a non-relativistic scalar in a generic 
NC background is:
\beq
\int d^{3} x \sqrt{g}
\left\{ i m v^\mu  \left(\phi^\dagger D_\mu \phi -
D_\mu \phi^\dagger \phi \right)  -h^{\mu \nu} D_\mu \phi^\dagger D_\nu \phi
-\xi R \phi^\dagger \phi
\right\} \, .
\label{sc1}
\eeq
Here the covariant derivative include just the gauge part:
\beq
D_\mu \phi = \p_\mu \phi - i m A_\mu \phi \, .
\eeq
One can obtain the action (\ref{sc1})
 from DLCQ reduction on a circle $x^-$ with radius $4 \pi$ 
of a relativistic scalar:
\beq
S=\frac{1}{4 \pi} \int d^{4} x \sqrt{-\det G_{AB}} \le 
-G^{MN} \p_M \Phi^\dagger \p_N \Phi  - \xi R \Phi^\dagger \Phi \ri  \, ,
\label{sc2}
\eeq
using
\beq
\Phi(x^-, x^\mu)=\phi(x^\mu) e^{i m x^-} \, .
\eeq

Let us specialize to the case with $A_\mu=0$.
We choose the positive branch $i \p_{t}=\sqrt{-\p_{t}^2}$
and we perform
 the Euclidean rotation:
$t \rightarrow -i t_E$, $m \rightarrow i m_E$;
we will omit the subscript $E$ in what follows.
In curved space, this can be realized by
\beq
v^\mu \arr i v^\mu \, ,  \qquad
 m \arr i m \, ,   \qquad 
n_\mu \arr -i n_\mu \, , \qquad
\sqrt{g} \arr i \sqrt{g}  \, .
\label{eu}
\eeq

Introducing the spatial laplacian:
\beq
\mathcal{D}^2 \phi=\frac{ \p_\mu ( \sqrt{g} h^{\mu \nu } \p_\nu \phi) }{\sqrt{g}} \, , 
\label{lapospaziale}
\eeq
we can write the euclidean action as:
\bea
S_E &= & - \int d^{3} x \sqrt{g} \, 
\phi^\dagger \hat{\triangle} \phi 
= 
\nl
&=&
\int d^{3} x \sqrt{g}
\phi^\dagger
\left\{ 
  m   v^\mu   \le \sqrt{- \p_\mu^2}  \phi \ri
+  m   \frac{  \sqrt{- \p_\mu^2} \le  \sqrt{g} \, v^\mu \,  \phi \ri}{\sqrt{g}}
  -   \mathcal{D}^2 \phi
+\xi R  \phi
\right\} 
\, .
\label{triatloncurvo}
\eea
We will consider perturbations around the flat
NC spacetime. It is convenient to split
the Schr\"odinger operator in a flat part plus a perturbation:
\beq
\label{triatlon}
\triangle= -2m \sqrt{-\p_t^2} + \p_i^2 \, , \qquad 
\hat{\triangle}=\triangle + \delta \triangle \, .
\eeq

\section{The heat kernel}

\label{sec3}

\subsection{The flat space case}

The relativistic conformal anomaly for a scalar was computed
by numerous authors, see e.g. \cite {Christensen:1977jc,Brown:1976wc,Dowker:1976zf,Hawking:1976ja}.
A convenient way to compute it is by the heat kernel formalism;
 see \cite{Birrell:1982ix,Vassilevich:2003xt,Mukhanov:2007zz} for reviews.
In flat space, for a euclidean relativistic scalar in $d_R$ dimensions,
the heat kernel is as follows:
\beq
K_{\p_i^2}(s;x,y)=
\langle y | e^{s \p_i^2} | x \rangle
=\frac{1}{(4 \pi s)^{d_R/2} } \exp \le -\frac{(x-y)^2}{4 s} \ri
\eeq
where $d_R$ is the total number of dimensions.
We will denote by $\tilde{K}_{\mathcal{O}}$ 
the restriction to $x=y$ of the
heat kernel $K_{\mathcal{O}}$ of the operator
$\mathcal{O}$, e.g.:
\beq
\tilde{K}_{\p_i^2}(s)= \frac{1}{(4 \pi s)^{d_R/2} } \, .
\label{tracciata1}
\eeq

We can use the results in \cite{Solodukhin:2009sk}
for the heat kernel in the non-relativistic case.
The following Euclidean negative-definite operator is introduced:
\beq
\label{tria1}
\triangle= \triangle_t + \p_i^2 \, , \qquad \triangle_t = -2m \sqrt{-\p_t^2} \, ,
\eeq
One can use the parameterization in \cite{Solodukhin:2009sk}:
\beq
\label{tria2}
e^{-2 m s  \sqrt{-\p_t^2}  }
=\int_0^\infty d \s \frac{ms}{\sqrt{\pi}} 
\frac{1}{\s^{3/2}} e^{-\frac{s^2 m^2}{\s}} e^{-\s (-\p_t^2)} \, ,
\eeq
 to rewrite the time-dependent part of the heat kernel as
\beq
K_{\triangle_t}=
\langle t | e^{-2 m s \sqrt{-\p_t^2}} | t' \rangle=
\int_0^\infty d \sigma \frac{m}{2 \pi} \frac{s}{\s^2}
e^{ -\frac{4 s^2 m^2 +(t-t')^2}{4 \sigma} }=
\frac{ms}{2 \pi} \frac{1}{m^2 s^2 +\frac{(t-t')^2}{4}} \, .
\eeq
The total heat kernel reads (here $d$ is the number of spatial
dimensions):
\beq
K_\triangle(s)=\langle x  t | e^{s \triangle} | y t' \rangle=
\frac{1}{2 \pi} \, \frac{ms}{m^2 s^2 +\frac{(t-t')^2}{4}}  \,
\frac{1}{(4 \pi s)^{d/2} } \exp \le -\frac{(x-y)^2}{4 s} \ri \, .
\eeq
The $(x,t)=(y,t')$ restriction of the unperturbed heat kernel is then:
\beq
\tilde{K}_{\triangle}(s)=
\langle x t | e^{s \triangle} | x t \rangle=
\frac{2}{m (4 \pi s)^{1+d/2} } \, .
\label{tracciata2}
\eeq
A comparison between eq.~(\ref{tracciata1})
and eq.~(\ref{tracciata2})
shows that the Schr\"odinger operator in $d+1$
spacetime dimensions feels the same spectral dimension
\beq
d_{\mathcal{O}}=-2 \frac{\p \log \tilde{K}_{\mathcal{O}}(s)}{\p \log s} \, ,
\eeq
as a relativistic laplacian in $d+2$ dimensions.
For this reason, the non-relativistic trace anomaly 
appears in odd spacetime dimensions and not in even ones 
as in the relativistic case.

\subsection{The curved-space heat kernel}

Let us consider the curved space correction for the heat kernel of the operator
$\hat{\triangle}$ defined in eq.~(\ref{triatloncurvo}).
One can decompose the total 
heat kernel as the sum of the flat space 
contribution generated by $\triangle$ and a correction 
due to $\delta \triangle$,
see eq.~(\ref{triatlon}). 
The operator
$\hat{\triangle}$ is usually defined as a differential operator
in the functional space with scalar product:
\beq
\langle x t | x' t' \rangle_{g} =\frac{\delta(x-x') \delta(t-t')}{\sqrt{g} } \, .
\eeq
One can define the heat kernel as
\beq
\hat{K}_{\hat{\triangle}}(s)=\exp (s \hat{\triangle} ) \, .
\eeq
The $(x,t)=(y,t')$ restriction of the heat kernel
 of the operator $\hat{\triangle}$ can be expanded 
in powers of $s$:
\beq
\tilde{K}_{\hat{\triangle}}(s)=
\langle x  t | e^{s \triangle} | x t \rangle_g=
\frac{1}{s^{d/2+1}} 
\le a_0(\hat{\triangle}) +a_2(\hat{\triangle}) s + a_4(\hat{\triangle}) 
s^2 + \dots \ri \, .
\eeq
We shall be interested in particular to the coefficient $a_4$,
which will give us the trace anomaly of the $d=2$ theory.

Let us sketch the derivation, taken  from \cite{Vassilevich:2003xt}.
A convenient expression for the renormalized vacuum functional can be given in terms of the $\zeta$ function 
\beq
W^{\rm ren}= -\frac{1}{2} \zeta'(0,\hat{\triangle})
 -\frac{1}{2} \log \mu^2 \zeta (0,\hat{\triangle}) \, ,
\eeq
where $\mu$ is a renormalization scale and the zeta function of the operator $\hat{\triangle}$
is defined as:
\beq
\zeta(s,\hat{\triangle})=\rm{ Tr}  ( {\hat{\triangle}}^{-s})    \, .
\eeq
Under a variation of the operator $\hat{\triangle}$,
the $\zeta$ function transforms as:
\beq
\delta \zeta(s,\hat{\triangle})=
-s \rm{Tr} ( (\delta \hat{\triangle}) \hat{\triangle}^{-s-1} ) \, . 
\eeq
Specializing to a Weyl transformation,
 the heat kernel generator transforms as:
\beq
\delta \hat{\triangle} =-2 \sigma \hat{\triangle} \, ,
\qquad
\delta \zeta=2 \s s \zeta \, .
\eeq
Moreover, the $\zeta$ function is regular at $s=0$.
Consequently, the variation of the vacuum functional is:
\beq
\delta W^{\rm ren} = -\zeta(0, \hat{\triangle})=-a_4(\hat{\triangle}) \, .
\label{a4zeta}
\eeq
The second equality in eq.~(\ref{a4zeta}) follows
from the relation:
\beq
\tilde{K}_{\hat{\triangle}}(t)=\frac{1}{2 \pi i} \oint ds \, t^{-s} \,
\Gamma(s) \zeta(s,\hat{\triangle})
\eeq
and by taking the residue at the pole at $s=0$.
This shows that for  $d=2$:
\beq
T^i_i-2 \epsilon^0 = a_4 (x, \hat{\triangle}) \, .
\eeq

It is useful to introduce an operator $\hat{M}$ for which
\beq
\langle x t | \hat{\triangle} | x' t' \rangle_g=
\langle x t | \hat{M} | x' t' \rangle \, , \qquad
{\rm where} \qquad
\langle x t | x' t' \rangle=\delta(x-x') \delta(t-t') \,.
\eeq
One can decompose the heat kernel
generator  $\hat{M}$ 
as the sum of the flat one plus a perturbation $\hat{V}$ that encodes all the gravitational effects:
\beq
\langle x t | \hat{M} | x' t' \rangle =
g^{1/4} ( \triangle + \delta{\triangle} ) [ g^{-1/4} \delta (x-x') \delta(t-t') ] \, ,
\qquad
\hat{M}=\triangle +\hat{V} \, . 
\eeq
where $\delta{\triangle}$ is the difference between the curved and 
the flat space Schr\"odinger operator,
see eq.~(\ref{triatlon}).
The trace of the operator $\hat{K}_{\hat{M}}$ can be expanded 
in powers of $s$:
\beq
\tilde{K}_{\hat{M}}(s)=
\langle x  t | e^{s \hat{M}} | x t \rangle
=\frac{1}{s^{d/2+1}} \le a_0(\hat{M}) +a_2(\hat{M}) s + a_4(\hat{M}) s^2 + \dots \ri \, .
\eeq
With our choice of conventions,
we have that
 $\tilde{K}_{\hat{M}}(s)=\sqrt{g} \, \tilde{K}_{\hat{\triangle}}(s)$.

\subsection{The metric perturbation}

To compute the curved space heat kernel, we specialize to a
simple perturbation in which just 
the background fields $(n_0, v^0)$ are perturbed in a time-independent way,
i.e. we take
\beq
h_{ij}=\delta_{ij} \, , \qquad n_i=v^i=0 \, , \qquad A_\mu=0 \, .
\eeq
We will use the following
parameterization:
\beq
n_0= \frac{1}{1-\eta(x)} \, , \qquad  v^0=1- \eta(x) \, ,
\qquad g^{1/2}= \frac{1}{1-\eta} \, ,  
\eeq
where $\eta(x)$ is a function of space.
We need $R$ at next-to-leading order:
\beq
R \approx -2 \p^2 \eta -2  \eta \p^2  \eta
-\frac{7}{2} \p_i  \eta \p_i  \eta +\dots \, .
\eeq
The dimension 4 curvature invariants are:
\bea
\label{leadcurv}
R^2 & \approx 4 & (\p^2  \eta)^2 \, , \qquad
 W^2 \approx \frac{1}{3} (\p^2  \eta)^2 
\, , \qquad 
E_4 \approx 2 ( \p^2  \eta)^2-2 \p_{ij}  \eta \p_{ij}  \eta \, ,
\nl
D_A D^A R & \approx &
-2 \p^2 \p^2 \eta
-2 (\p^2 \eta)^2 
-2 \eta \p^2 \p^2 \eta
-13\p_k \eta \p_k \p^2 \eta  
-7 \p_{ij}  \eta \p_{ij}  \eta \, .
\eea

The covariant spatial laplacian, as defined in eq.~(\ref{lapospaziale}), is:
\beq
\mathcal{D}^2 \phi =g^{-1/2} \p_i ({g}^{1/2} \p_i \phi ) 
\approx \p^2 \phi +( \p_i \eta +\eta \p_i \eta) \p_i \phi \, .
\eeq
One should also take into account the
 normalization factors
 of the $\delta$ function in the heat kernel, 
 which at the second order 
in $\eta$ reads:
\beq
-g^{1/4} \mathcal{D}^2 (g^{-1/4} \delta(x))
\approx
-\p^2 \delta(x)
+\delta(x) \le \frac{\p^2 \eta}{2} +\frac{1}{2} \eta \p^2 \eta +\frac{3}{4} \p_i \eta \p_i \eta \ri \, .
\eeq
In the Euclidean,
the heat kernel generator $\hat{M}$ is:
\beq
\langle x t | \hat{M} | x' t' \rangle=
\langle x t | \le  \triangle
+ S(x)  \sqrt{-\p_{0}^2} \delta(x-x')  \delta(t-t')
+ P(x) \delta(x-x') \delta(t-t')
\ri | x' t' \rangle \, ,
\eeq
where 
\beq 
S=2 m \eta \, , \qquad
P= - \le \frac{\p^2 \eta}{2} +\frac{1}{2} \eta \p^2 \eta +\frac{3}{4} \p_i \eta \p_i \eta \ri
+\xi
\le 2 \p^2 \eta + 2 \eta \p^2 \eta + \frac{7}{2} (\p_i \eta)^2 \ri
\, .
\label{SP}
\eeq

\subsection{The perturbative calculation}

One can use the perturbative approach explained for example in the textbook
\cite{Mukhanov:2007zz} and in the paper \cite{Barvinsky:1990up}:
\beq
K_{\hat{M}} (s)=\exp(s (\triangle+ \hV)) =\sum_{n=0}^{\infty} K_n(s) \, ,
\eeq
where
\beq
K_n(s)=\int_{0}^{s} d s_n \int_{0}^{s_n} ds_{n-1} \dots \int_0^{s_2} ds_1
e^{(s-s_n)\triangle} \hV e^{(s_n-s_{n-1}) \triangle} \hV \dots
e^{(s_2-s_1) \triangle} \hV e^{s_1 \triangle} \, .
\eeq
The operator $\hat{V}$ is given by:
\beq
\langle x t | \hat{V} | x' t' \rangle=
\langle x t | \le 
 S(x)  \sqrt{-\p_{0}^2} \delta(x-x') \delta(t-t')
+ P(x) \delta(x-x') \delta(t-t')
\ri | x' t' \rangle \, .
\eeq
To determine $a_4$ at the lowest order in $\eta$,
we need to compute the $K_1$ contribution using
the functions $P,S$ in eq.~(\ref{SP}) at the second order in $\eta$
and the $K_2$ contribution using $P,S$  at the first order in $\eta$.

The contribution from $K_1$ splits in a part due to $P$ and a part due to $S$:
\bea
\tilde{K}_{1P} & = &
 \frac{2}{m(4 \pi s )^{d/2+1} } 
\le s P +\frac{1}{6} s^2 \p^2_x P + \dots \ri
\label{primoP}
\nl
\tilde{K}_{1S} & = &
  \frac{ 1 }{m^2} 
  \frac{1}{(4 \pi s )^{d/2+1} }\
\le  S +\frac{s}{6} \p_x^2 S
+\frac{s^2}{60} \p^2 \p^2 S
+\dots\ri \, .
\label{primoS}
\eea
The contribution due to $K_2$ splits in four pieces:
\bea
\label{secondiaiosa}
\tilde{K}_{2SS} & = &
 \frac{1}{2 m^3 (4 \pi s)^{d/2+1}}  \le  S^2 
 +\frac{s}{3} S \p_k^2 S + \frac{s}{6} \p_k S \p_k S 
\right.
\nl & &
\left. 
 + \frac{s^2}{30}   S \p^2 \p^2 S 
+\frac{s^2}{36} \p^2 S \p^2 S
+\frac{s^2}{15} \p_i \p^2 S \p_i S
+\frac{s^2}{45} \p_i \p_j S \p_i \p_j S
+\dots 
\right)
\nl
\tilde{K}_{2PP}
& = & \frac{2}{m  (4 \pi s)^{d/2+1}} \le \frac{s^2}{2} P(x)^2 + \dots \ri
\nl
\tilde{K}_{2PS} & = & K_{2SP}=
\frac{1}{m^2 (4 \pi s)^{d/2+1} } 
\le \frac{s}{2} S P +\frac{s^2}{12} \le \p^2 S P + S \p^2 P 
+ \p_i S \p_i P\ri + \dots\ri \, .
\eea
The calculation are sketched
in appendices \ref{AppeA} and \ref{AppeB}.

We can then re-express the $a_4$ coefficient in terms
of the curvature invariants, see eq.~(\ref{leadcurv}).
There is a degeneracy between  $W^2$ and  $R^2$, 
due to the fact that in the simple background that we have chosen
they are proportional to each other. In order to fix these coefficients, we can
use the fact that for the conformal coupling $\xi=1/6$, the coefficient of the $R^2$
term must vanish due to the Wess-Zumino consistency conditions.
Up to quadratic order in $\eta$, the result is:
\beq
a_4(\hat{M}) = \sqrt{g}
\left( -a E_4 + c W^2 + b R^2 + d D_A D^A R  \right) \, ,
\eeq
where the coefficients are
 given in eq.~(\ref{mainresult2}).

\section{Conclusions}

\label{sec4}

It is natural to conjecture that an analogous of the $a$
theorem may hold for the $E_4$ anomaly coefficient
 in $d=2$ Schr\"odinger-invariant  theories.
If we consider two fixed
points in the UV and in the IR with matter content given
just by free scalars, it would mean that
the following quantity should decrease from UV to IR:
\beq
a_{UV} \propto \sum_k^{UV} \frac{1}{m_k} \geq
\sum_k^{IR} \frac{1}{m_k}
\propto a_{IR} \, ,
\label{aaath}
\eeq
where the sum over $k$ is over the number of scalar species.
In this class of theories indeed the mass is conserved,
and so the mass of bound states is the sum of the elementary
constituents, with no bound-state deficit mass. 
The statement in eq.~(\ref{aaath})
may give a quantitative formulation to the physical intuition
that bound states should form in the IR:
in the process of adding 
energy to a system,
bound states are broken instead of formed.

Up to an  overall $1/m$ factor, 
 the anomaly coefficients
 in eq.~(\ref{mainresult2}) are numerically identical 
 to the  corresponding ones in the relativistic case in 4 dimensions,
 see e.g. \cite {Christensen:1977jc,Brown:1976wc,Dowker:1976zf,Hawking:1976ja}.
 It would be interesting to check if this 
 numerical coincidence is valid in more general cases
 (e.g. for fermions)
 and if it has a physical explanation. 
 
 Several aspects deserve consideration for further investigations,
 for instance the trace anomaly
 for a free fermion. 
 The Chern-Simons term is also very interesting: 
 it describes anyons, and in three spacetime dimension
 this gives a continuous interpolation between the 
 bosonic and the fermionic case.
 These calculations will be useful to check the conjectured
 $a$-theorem in practical condensed-matter examples. 
Possible techniques which might be used for
a proof are the local renormalization group 
\cite{Osborn:1989td,Jack:1990eb,Osborn:1991gm}
and the dispersion relations method
\cite{Komargodski:2011vj,Komargodski:2011xv}.

We considered only the simplest sector $N_n=0$
in the anomaly, while an infinite number of higher derivatives
terms is present in the other sectors, e.g. eq.~(\ref{anoma-esempio}).
The general structure of the anomaly terms with 
an arbitrary number of $n_A$ is not known and in particular
it is not known if additional type A anomalies are present.

In the supersymmetric case,
it is possible that exact expressions
for $a$ might be found also in the interacting case,
by coupling the theory to a supergravity background
as in \cite{Anselmi:1997am,Intriligator:2003jj,Auzzi:2015yia}.
Newton-Cartan supergravity was recently 
studied in  \cite{Bergshoeff:2015uaa,Bergshoeff:2015ija}.
Another interesting direction is holography
\cite{Son:2008ye,Balasubramanian:2008dm,Maldacena:2008wh,Myers:2010tj,Liu:2015xxa,Taylor:2015glc}.

\section*{Acknowledgments}

We are grateful to Carlos Hoyos for useful comments.

\section*{Appendix}
\addtocontents{toc}{\protect\setcounter{tocdepth}{1}}
\appendix

\section{Heat kernel at first order}

\label{AppeA}

Let us first consider a multiplicative perturbation $P(x)$:
\bea
 K_{1P}(s) & = &
  \int_0^s ds' \int d^d \tilde{x} \int d \tilde{t}
\langle x t | e^{- (s-s') \triangle} | \tilde{x} \tilde{t} \rangle
 P(\tilde{x})
  \langle  \tilde{x} \tilde{t} | e^{-s' \triangle}| y  t' \rangle
\nl
& = &
\frac{ 1}{(2 \pi)^2}  
  \int_0^s ds'
   \frac{1}{(4 \pi (s-s'))^{d/2} } \,
 \frac{1}{(4 \pi s')^{d/2} } 
  \int d \tilde{t}
\frac{m(s-s')}{m^2 (s-s')^2 +\frac{(t-\tilde{t})^2}{4}} \, 
 \frac{ms'}{m^2 s'^2 +\frac{(\tilde{t}-t')^2}{4}}
\nl 
& &
    \int d^d \tilde{x}
 \, P(\tilde{x})
 \exp \le
 -\frac{(x-\tilde{x})^2}{4 (s-s')} -\frac{(\tilde{x}-y)^2}{4 s'} \ri \, .
\eea
The $d\tilde{t}$ integral can be computed explicitly; moreover we
can Fourier transform $P$:
\beq
P (\tilde{x})=\int \frac{d^d k}{(2 \pi)^{d/2}} e^{i k \tilde{x}} P(k) \, .
\eeq
We get the following expression:
\bea
 K_{1P}(s) & = &
\frac{ 1}{(2 \pi)^2}  
  \int_0^s ds'
   \frac{1}{(4 \pi (s-s'))^{d/2} } \,
 \frac{1}{(4 \pi s')^{d/2} }  \,
\frac{8 m \pi s}{ 4 m^2 s^2 + (t-t')^2}
\nl &  &
    \int d^d \tilde{x}
 \,  \int \frac{d^d k}{(2 \pi)^{d/2}} P(k)
 \exp \le
 -\frac{(x-\tilde{x})^2}{4 (s-s')} -\frac{(\tilde{x}-y)^2}{4 s'} 
+i k \tilde{x} \ri \, .
\eea
Doing the gaussian integral, we get:
\bea
& &
 K_{1P}(s) =
\frac{ 1}{(2 \pi)^2}  
  \int_0^s ds'
 \frac{1}{(4 \pi s )^{d/2} } \,
\frac{8 m \pi s}{ 4 m^2 s^2 + (t-t')^2}
\label{primoP2}
\nl & &
\int \frac{d^d k}{2 \pi^{d/2}}  \, 
\exp \le -\frac{(x-y)^2}{4 s} + ik \cdot \le x \frac{s'}{s} 
+y \frac{s-s'}{s} \ri -k^2 \frac{s'}{s} (s-s')\ri
P(k) \, .
\eea
Setting $t=t'$ and $x=y$ and expanding we recover
 the first of eq.~(\ref{primoP}).

The single insertion of $S$ can be reduces to 
a derivative acting on the single insertion of a $P$:
\bea
K_{1S}(s) &=&
  \int_0^s ds' \int d^d \tilde{x} \int d \tilde{t}
\langle x t | e^{- (s-s') \triangle} | \tilde{x} \tilde{t} \rangle
 S(\tilde{x})
 \sqrt{-\p_{\tilde{t}}^2}   
  \langle  \tilde{x} \tilde{t} | e^{-s' \triangle}| y t' \rangle
\nl & = &  
   \sqrt{-\p_{{t'}}^2}  \le
  \int_0^s ds' \int d^d \tilde{x} \int d \tilde{t}
\langle x t | e^{- (s-s') \triangle} | \tilde{x} \tilde{t} \rangle
 S(\tilde{x})
  \langle  \tilde{x} \tilde{t} | e^{-s' \triangle}| y t' \rangle
  \ri \, .
\eea
We can now use eq.~(\ref{primoP2}).
In order to perform the $ \sqrt{-\p_{{t'}}^2} $ operator,
we can use the following formula:
\beq
\sqrt{-\p_{{t}}^2} \le
 \frac{1}{ 1+\frac{t^2}{A^2}}
 \ri = \frac{A (A^2 -t^2)}{(A^2 +t^2)^2} \, ,
  \label{dederi2}
\eeq
which can be derived directly using Fourier transform:
\beq
\mathcal{F}
\le\frac{1}{ 1+\frac{t^2}{A^2}}\ri=\sqrt{\frac{\pi}{2}} A e^{-A |\omega|}
\, .
\eeq

\section{Heat kernel at second order}

\label{AppeB}

The four contributions  $K_{2 X_1 X_2}(s)$, where
\beq
X_1=\left\{ P(x_1), S(x_1)  \right\} \, , 
\qquad
X_2=\left\{ P(x_2), S (x_2)  \right\} \, , 
\eeq
 have a very similar structure:
\beq
K_{2 X_1 X_2}(s)=
\int_0^s d s_2 \int_0^{s_2} ds_1
\langle x' t' | e^{-(s-s_2)\triangle}
| x_2 t_2 \rangle 
\hat{X}_2
\langle x_2 t_2 | e^{-(s_2-s_1) \triangle}  | x_1 t_1 \rangle
\hat{X}_1
\langle 
x_1 t_1 | e^{-s_1 \Delta }
| x t \rangle \, ,
\eeq
where
\beq
\hat{X}_1=\left\{ P(x_1), S(x_1)  \sqrt{- \p_{t_1}^2} \right\} \, , 
\qquad
\hat{X}_2=\left\{ P(x_2), S (x_2) \sqrt{- \p_{t_2}^2} \right\} \, . 
\eeq

We can split it as follows:
\beq
K_{2 X_1 X_2}(s)=\int_0^s d s_2 \int_0^{s_2} ds_1
\frac{1}{(4 \pi (s-s_2))^{d/2} } \, 
\frac{1}{(4 \pi (s_2-s_1))^{d/2} }  \, 
\frac{1}{(4 \pi s_1)^{d/2} } \Xi^{X_1 X_2} \,  \Psi^{X_1 X_2} \, ,
\eeq
where $\Xi^{X_1 X_2}$ and $\Psi^{X_1 X_2}$
correspond to the space and time part of the integrals.
The space part is:
\begin{eqnarray*}
 \Xi^{X_1 X_2} &=&
\int d x_1 \int d x_2 
\exp \le  i k_1 x_1+i k_2 x_2 
 -\frac{(x'-x_2)^2}{4 (s-s_2)}
 -\frac{(x_2-x_1)^2}{4 (s_2-s_1)} 
-\frac{(x_1-x)^2}{4 s_1}\ri
 X_1(k_1) X_2(k_2)   
\nl
&=& (4 \pi)^{d} \le \frac{s_1 (s-s_2) (s_2-s_1)}{s} \ri^{d/2}
\exp \le 
\frac{i k_1 s_1 x'}{s}+\frac{i k_2 s_2 x'}{s}-\frac{i k_1 s_1 x}{s}
-\frac{i k_2 s_2 x}{s}+\frac{k_1^2 s_1^2}{s}+\frac{k_2^2 s_2^2}{s}
\right.
\nl & &
\left.
-k_1^2   s_1-2 k_1 k_2 s_1-k_2^2 s_2+\frac{2 k_1 k_2 s_1 s_2}{s}+i k_1 x+i k_2 x
-\frac{x^2}{4 s}+\frac{x x'}{2 s}-\frac{\left(x'\right)^2}{4 s}
\ri X_1 X_2  \, ,
\end{eqnarray*}
which, specializing for $x=x'$, reads:
\bea
 \Xi^{X_1 X_2}|_{x=x'} &=&
 \exp \le 
- \le \frac{s_1^2}{s} -s_1 \ri \p^2_{x_1}
- \le \frac{s_2^2}{s} -s_2 \ri \p^2_{x_2}
 - 2  \le \frac{s_1 s_2}{s} -s_1 \ri \p_{x_1} \cdot \p_{x_2}
\ri X_1(x_1) X_2(x_2) 
\nl
& &
(4 \pi)^{d} \le \frac{s_1 (s-s_2) (s_2-s_1)}{s} \ri^{d/2}
\, .
\eea
The time part is:
 \begin{eqnarray*}
\Psi^{PP} &=&
\frac{1}{(2  \pi)^3}
 \int \! d t_1  \int \!  d t_2
 \frac{m (s-s_2)}{m^2 (s-s_2)^2 +\frac{(t_2-t')^2}{4}} \,
 \frac{m (s_2-s_1)}{m^2 (s_2-s_1)^2 +\frac{(t_2-t_1)^2}{4}} \,
 \frac{ms_1}{m^2 s_1^2 +\frac{(t_1-t)^2}{4}} \, ,
\nl
\Psi^{SP} &=&
\frac{1}{4  \pi^3}
 \int \! d t_1  \int \!  d t_2
 \frac{m (s-s_2)}{m^2 (s-s_2)^2 +\frac{(t_2-t')^2}{4}} \, 
 \frac{m (s_2-s_1)}{m^2 (s_2-s_1)^2 +\frac{(t_2-t_1)^2}{4}} \, 
 \frac{ 4m^2 s_1^2 -(t_1-t)^2}{(4 m^2 s_1^2 +(t_1-t)^2)^2} \, ,
\nl
\Psi^{PS} &=&
\frac{1}{4  \pi^3}
 \int \! d t_1  \int \!  d t_2
 \frac{m (s-s_2)}{m^2 (s-s_2)^2 +\frac{(t_2-t')^2}{4}} \, 
 \frac{ 4m^2 (s_2-s_1)^2 -(t_2-t_1)^2}{(4 m^2 (s_2-s_1)^2 +(t_2-t_1)^2)^2} \,
 \frac{ms_1}{m^2 s_1^2 +\frac{(t_1-t)^2}{4}} \, ,
\nl
\Psi^{SS} &=&
\frac{1}{2  \pi^3}
  \int \! d t_1  \int \!  d t_2
 \frac{m (s-s_2)}{m^2 (s-s_2)^2 +\frac{(t_2-t')^2}{4}}  \, 
 \frac{ 4m^2 (s_2-s_1)^2 -(t_2-t_1)^2}{(4 m^2 (s_2-s_1)^2 +(t_2-t_1)^2)^2}  \,
 \frac{ 4m^2 s_1^2 -(t_1-t)^2}{(4 m^2 s_1^2 +(t_1-t)^2)^2} \, .
\end{eqnarray*}
The result of the integration is:
\bea
\Psi^{PP}&=&\frac{1}{\pi} \frac{2 m s}{4 m^2 s^2+\left(t-t'\right)^2} \, ,
\qquad
\Psi^{SS}=\frac{1}{\pi}
\frac{4  m s \left(4 m^2 s^2-3 \left(t-t'\right)^2\right)}{\left(4 m^2 s^2+\left(t-t'\right)^2\right)^3} \, ,
\nl
\Psi^{PS}&=&\Psi^{SP}=
\frac{1}{ \pi}
\frac{\left(4 m^2 s^2-\left(t-t'\right)^2\right)}{\left(4 m^2 s^2+\left(t-t'\right)^2\right)^2} \, .
 \eea
Putting all together and specializing to $t=t'$ and $x=x'$,
we find the expressions in eq.~(\ref{secondiaiosa}).


\end{document}